\documentclass[%
 reprint,
 amsmath,amssymb,
 nrc1,
]{revtex4-1}

\usepackage{nrc1}
\usepackage{graphicx}% Include figure files
\usepackage{dcolumn}% Align table columns on decimal point
\usepackage{bm}% bold math
\usepackage[hypertex, linktocpage]{hyperref}[2003/11/30]
\usepackage{siunitx}
\usepackage{bm}
\usepackage{verbatim}

\begin{document}

\preprint{APS/123-QED}

\title{Probing Split-Ring Resonator Permeabilities with Loop-Gap Resonators}
%\title[Sample title]{Sample Title:\\with Forced Linebreak\footnote{Error!}}% Force line breaks with \\
%\thanks{Footnote to title of article.}

\author{J. S. Bobowski}
\affiliation{Department of Physics, University of British Columbia, Kelowna, British Columbia, V1V 1V7, Canada}
\email{jake.bobowski@ubc.ca}
\homepage{\\https://people.ok.ubc.ca/jbobowsk}
%\author{A. Author}
% \altaffiliation[Also at ]{Physics Department, XYZ University.}%Lines break automatically or can be forced with \\
%\author{B. Author}%
% \email{Second.Author@institution.edu.}
%\affiliation{ 
%Authors' institution and/or address%\\This line break forced with \textbackslash\textbackslash
%}%

%\author{C. Author}
% \homepage{http://www.Second.institution.edu/~Charlie.Author.}
%\affiliation{%
%Second institution and/or address%\\This line break forced% with \\
%}%

\date{\today}% It is always \today, today,
             %  but any date may be explicitly specified

\begin{abstract}
A method is proposed to experimentally determine the effective complex permeability of split-ring resonator (SRR) arrays used in the design of metamaterials at microwave frequencies.  We analyze the microwave response of a loop-gap resonator (LGR) whose bore has been partially loaded with one or more SRRs.  Our analysis reveals that the resonance frequency, magnetic plasma frequency, and damping constant of the effective permeability of the SRR array can be extracted from fits to the reflection coefficient ($S_{11}$) of an inductively-coupled LGR.  We propose LGR designs that would allow both a one-dimensional array of SRRs and small three-dimensional arrays of SRRs to be characterized.  Finally, we demonstrate the method using a toroidal LGR loaded with a single extended SRR of length $z$.
%
%Valid PACS numbers may be entered using the \verb+\pacs{#1}+ command.
\end{abstract}

%\pacs{Valid PACS appear here}% PACS, the Physics and Astronomy
                             % Classification Scheme.
%\keywords{Suggested keywords}%Use showkeys class option if keyword
                              %display desired
\maketitle

%\begin{quotation}
%The ``lead paragraph'' is encapsulated with the \LaTeX\ 
%\verb+quotation+ environment and is formatted as a single paragraph before the first section heading. 
%(The \verb+quotation+ environment reverts to its usual meaning after the first sectioning command.) 
%Note that numbered references are allowed in the lead paragraph.
%
%The lead paragraph will only be found in an article being prepared for the journal \textit{Chaos}.
%\end{quotation}

%\section{\label{sec:intro}First-level heading:\protect\\ The line break was forced \lowercase{via} \textbackslash\textbackslash}

\section{Introduction}
In 1981, Hardy and Whitehead described a high-$Q$ rf/microwave resonator built from a conducting tube with a narrow slit along its length.  The advantages of this type of resonator, which they termed a split-ring resonator (SRR), include design flexibility, simple and inexpensive construction, good isolation between electric and magnetic fields, and uniform field distributions~\cite{Hardy:1981}.  Froncisz and Hyde were quick to adapt this resonator design for the purposes of electron spin resonance (ESR) experiments~\cite{Froncisz:1982}.  In this work, Froncisz and Hyde also chose to rename the resonator a loop-gap resonator (LGR).  Since then, these authors have continued to develop these resonators for ESR applications and the term LGR has largely prevailed.  Reviews of this body of work are given in Refs.~[\onlinecite{Hyde:1986, Hyde:1989, Rinard:2005}].  It is worth noting, however, that these resonators have, in some cases, continued to be referred to as SRRs in the literature~\cite{Hall:1985, Bonn:1991, Hardy:1993, Bobowski:2013, Bobowski:2015, Bobowski:2017}.

In 1999, Pendry {\it et al.\@} proposed methods for engineering metamaterials with negative permeability over a narrow range of frequencies in the microwave regime~\cite{Pendry:1999}.  Each of the methods make use of an array of resonant structures. In one case, each element of the array consists of a pair of concentric ``split-ring'' cylinders made from a conducting foil (see Figs.~2 and 3 in Ref.~[\onlinecite{Pendry:1999}])~\cite{Ghim:1996}.  Pendry {\it et al.\@} showed that these split-ring structures could be engineered to have negative effective permeabilities at frequencies just above their resonant frequency~\cite{Pendry:1999}.

Furthermore, these authors proposed that a 1-D array of disk-like planar split-rings, each of negligible length, could be substituted for the concentric split cylinders (see Fig.~12 in Ref.~[\onlinecite{Pendry:1999}]).  Provided that the spacing of the disks is less than the ring radius, the behaviour of the 1-D array could be made to mimic that of the split cylinders for microwave magnetic fields applied parallel to the ring/cylinder axes.  These planar structures are now widely known as split-ring resonators (SRRs).  Negative permeability in metamaterials fabricated from arrays of SRRs has since been experimentally confirmed~\cite{Smith:2000}.  SRRs have also been used in microwave metamaterials that exhibit a negative refraction index, i.e.\@ materials with both an effective permittivity and permeability that are simultaneously less than zero~\cite{Shelby:2001}.

In this paper, resonators fabricated from bulk conductors are called LGRs, concentric split cylinders of length $z$ made using metallic foils are called extended SRRs (ESRRs), and planar split-ring structures are called SRRs.  

Recently, we have been focused on using LGRs to characterize the electromagnetic (EM) properties of various materials~\cite{Bobowski:2013, Bobowski:2015, Bobowski:2017, Dubreuil:2017}. Building on this body of work, we now propose a method that allows one to experimentally determine the effective complex permeability of either a 1-D array or a small 3-D array of SRRs.  In our method, the bore of a LGR of length $\ell$ is partially loaded with SRRs that occupy a length $z$, with the restriction that $z<\ell$.  The presence of the SRR array, with an effective relative permeability \mbox{$\mu_\mathrm{r}=\mu^\prime - j\mu^{\prime\prime}$}, modifies the inductance $L$ of the LGR, thereby changing its resonant response.  Fits to the measured response can then be used to extract the parameters (resonant frequency, plasma frequency, and damping constant) that characterize the permeability of the SRR array.

This work has, in part, been motivated by the fact that extracting the complex permittivity and/or permeability of microwave metamaterials has proven to be challenging.  These effective parameters are typically extracted using a retrieval technique that relies on scattering parameters obtained from numerical simulations of the metamaterial structure~\cite{Smith:2002}.  Using these methods, several authors have reported the surprising result that the imaginary component of either the permittivity or permeability of a particular metamaterial is negative over a narrow band of frequencies, a so-called antiresonant response~\cite{OBrien:2002, Koschny:2003, Markos:2003, Seetharamdoo:2005}.  

Woodley and Mojahedi have since argued that the Lorentzian model, widely used to characterize the effective parameters of metamaterials, does not allow for negative imaginary components.  Furthermore, these authors demonstrated that the retrieval technique, applied to an array of dielectric spheres, resulted in an imaginary permittivity that is negative for a range of frequencies~\cite{Woodley:2010}.  However, the effective parameters for this structure can be solved for analytically and it is known that the imaginary components of both the permittivity and permeability are positive for all frequencies~\cite{Wheeler:2005}.  Therefore, Woodley and Mojahedi have posited that the reported negative values of the imaginary components of metamaterial effective parameters are the result of numerical errors in the simulations~\cite{Woodley:2010}.  

Our work proposes an experimental technique that can be used to determine the complex permeability of a small array of resonant elements used to construct metamaterials at microwave frequencies. The outline of this paper is as follows: In Sec.~\ref{sec:SRRLGR}, the microwave response of a LGR whose bore has been partially filled with a 1-D SRR array is calculated.  We show that a measurement of this response allows the complex permeability of the SRR array to be experimentally probed.  In Sec.~\ref{sec:multiple}, we show how this method can be adapted to accommodate small 3-D arrays of SRRs within the bore of a multi-loop, multi-gap LGR.  In Sec.~\ref{sec:expt}, we demonstrate the experimental method using a toroidal LGR (TLGR) that has been loaded with a single ESRR.  Section~\ref{sec:negMu2} examines the experimental signature that would be expected from a LGR loaded with a material having a negative imaginary permeability.  Finally, the main conclusions are summarized in Sec.~\ref{sec:summary}.

\section{Microwave Response of a SRR-loaded LGR}\label{sec:SRRLGR}
\subsection{LGR design equations}
\begin{figure}
\includegraphics[width=4 cm]{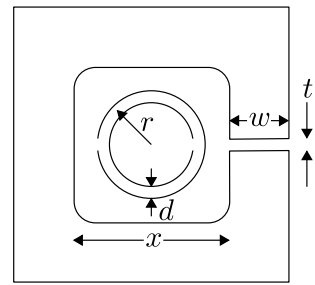}% Here is how to import EPS art
\caption{\label{fig:1loop1gap}A cross-sectional view of a LGR loaded with a singe ESRR.  The bore of the LGR has a square cross-section of with sides of length $x$, the gap width is $w$, and the gap height is $t$.  The ESRR has an average radius $r$ and the spacing between the concentric rings is $d$.  For a LGR loaded with an ESRR or a 1-D SRR array, $x=a$ where $a$ defines the size of the SRR unit cell.}
\end{figure}

Figure~\ref{fig:1loop1gap} shows a cross-sectional view of a rectangular LGR loaded with a single ESRR.  The general methods presented, however, can also be applied to a 1-D array of SRRs.  The critical LGR and ESRR dimensions are labelled in the figure.  
The corners of the LGR bore have been drawn with a radius so as to avoid the high current densities that would exist at the sharp corners of a perfect square.  For a LGR of length $\ell$, the gap capacitance and loop inductance are given by \mbox{$C_0\approx \varepsilon_0 w\ell/t$} and \mbox{$L_0\approx \mu_0 x^2/\ell$}, respectively.  It follows that the resonant frequency of the LGR, when its gap and bore are empty, is approximately given by
\begin{equation}
\omega_0=\frac{1}{\sqrt{L_0C_0}}\approx\frac{c}{x}\sqrt{\frac{t}{w}},\label{eq:w0}
\end{equation}
where $c=1/\sqrt{\mu_0\varepsilon_0}$ is the vacuum speed of light.  The expression above ignores the effects of fringing electric and magnetic fields.  For more accurate design equations, see Ref.~[\onlinecite{Rinard:2005}] and the references therein.  We note, however, that the resonant frequency and quality factor (discussed below) of the empty LGR can be accurately determined experimentally.  Therefore, for the purposes of this work, the approximate design equations are adequate.

If an external shield is used to suppress radiative losses, the quality factor of an empty LGR is given by
\begin{equation}
Q_0=\frac{1}{R_0}\sqrt{\frac{L_0}{C_0}}\approx\frac{x}{2\delta},
\end{equation}
where $R_0$ is the effective resistance of the LGR at $\omega_0$, \mbox{$\delta=\sqrt{2\rho/\left(\mu_0\omega\right)}$} is the EM skin depth, and $\rho$ is the resistivity of the conductor used to construct the LGR~\cite{Hardy:1981, Bobowski:2013}.

\subsection{SRR design equations}
We first note that, to be consistent with our previous work, we have adopted the sign convention \mbox{$\mu_\mathrm{r}=\mu^\prime -j\mu^{\prime\prime}$}~\cite{Bobowski:2015, Bobowski:2017, Dubreuil:2017}.  Written in terms of the SRR resonant frequency $\omega_\mathrm{S}$, plasma frequency $\omega_\mathrm{P}$, and damping constant $\gamma$, the real and imaginary parts of the SRR relative permeability, first calculated by Pendry {\it et al.}, are given by~\cite{Pendry:1999}
\begin{align}
\mu^\prime &=1-\dfrac{\left[1-\left(\dfrac{\omega_\mathrm{S}}{\omega_\mathrm{P}}\right)^2\right]\left[1-\left(\dfrac{\omega_\mathrm{S}}{\omega}\right)^2\right]}{\left[1-\left(\dfrac{\omega_\mathrm{S}}{\omega}\right)^2\right]^2+\left(\dfrac{\gamma}{\omega}\right)^2},\label{eq:mu1}\\
\mu^{\prime\prime} &=\dfrac{\dfrac{\gamma}{\omega}\left[1-\left(\dfrac{\omega_\mathrm{S}}{\omega_\mathrm{P}}\right)^2\right]}{\left[1-\left(\dfrac{\omega_\mathrm{S}}{\omega}\right)^2\right]^2+\left(\dfrac{\gamma}{\omega}\right)^2},\label{eq:mu2}
\end{align}
where $\omega$ is angular frequency.  Equations (\ref{eq:mu1}) and (\ref{eq:mu2}) are valid for both ESRRs and SRR arrays.  In terms of the ESRR dimensions shown in Fig.~\ref{fig:1loop1gap}
\begin{align}
\omega_\mathrm{S} &=\sqrt{\frac{3dc^2}{\pi^2r^3}},\\
\omega_\mathrm{P} &=\frac{\omega_\mathrm{S}}{\sqrt{F}},\label{eq:wP}\\
\gamma &=\frac{2\sigma}{\mu_0 r}.
\end{align}
In the above expressions \mbox{$F=1-\pi r^2/a^2$} and $\sigma=\rho/\delta$ is the sheet resistance of the metallic foil used to form the ESRR.  When one is dealing with an array of SRRs, $a$ represents the size of a unit cell.  For a single ESRR loaded in a LGR, as in Fig.~\ref{fig:1loop1gap}, $a=x$.  Strictly speaking, $\gamma$ is frequency dependent since \mbox{$\delta\propto \omega^{-1/2}$}.  In Eqs.~(\ref{eq:mu1}) and (\ref{eq:mu2}), it is convenient to re-express the factor $\gamma/\omega$ as
\begin{equation}
\frac{\gamma}{\omega} = \frac{\delta}{r} = \frac{\delta_\mathrm{S}}{r}\sqrt{\frac{\omega_\mathrm{S}}{\omega}} = \frac{\gamma_\mathrm{S}}{\omega}\sqrt{\frac{\omega_\mathrm{S}}{\omega}},\label{eq:gammaS}
\end{equation}
where $\delta_\mathrm{S}$ and $\gamma_\mathrm{S}$ are the skin depth and damping constant evaluated at the resonant frequency of the ESRR, respectively.

\subsection{Impedance of an empty LGR}
\begin{figure}
\includegraphics[width=6 cm]{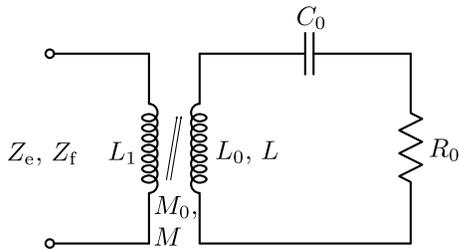}% Here is how to import EPS art
\caption{\label{fig:mutual}An equivalent circuit model of a one-loop, one-gap LGR with the bore of the resonator inductively coupled to an inductance $L_1$.  The mutual inductance between $L_1$ and an empty (partially-filled) LGR of inductance $L_0$ $\left(L\right)$ is $M_0$ $\left(M\right)$.}
\end{figure}

Although several methods exist for coupling signals into and out of the LGR, the most widely used method makes use of a length of coaxial transmission line shorted by a wire loop that couples magnetic flux into and out of the bore of the resonator~\cite{Rinard:1993}.  If the coupling loop has inductance $L_1$ and the mutual inductance between the coupling loop and the LGR is $M_0$, then the equivalent circuit of the coupled resonator is as shown in Fig.~\ref{fig:mutual}.  

Rinard {\it et al.\@} have solved for the impedance of an inductively coupled LGR (see Eq.~(22) in Ref.~[\onlinecite{Rinard:1993}]).  Rewritten in terms of the LGR resonant frequency $\omega_0$ and quality factor $Q_0$, the impedance $Z_\mathrm{e}=R_\mathrm{e}+jX_\mathrm{e}$ is given by
\begin{align}
R_\mathrm{e} &=\frac{\dfrac{\left(\omega M_0\right)^2}{R_0}\sqrt{\dfrac{\omega}{\omega_0}}}{\dfrac{\omega}{\omega_0}+Q_0^2\left(\dfrac{\omega}{\omega_0}-\dfrac{\omega_0}{\omega}\right)^2},\label{eq:Re}\\
X_\mathrm{e} &=\omega L_1-\frac{\dfrac{\left(\omega M_0\right)^2}{R_0}Q_0\left(\dfrac{\omega}{\omega_0}-\dfrac{\omega_0}{\omega}\right)}{\dfrac{\omega}{\omega_0}+Q_0^2\left(\dfrac{\omega}{\omega_0}-\dfrac{\omega_0}{\omega}\right)^2}.\label{eq:Xe}
\end{align}
The subscript ``e'' serves as a reminder that these expressions are only valid when the bore of the LGR is empty.  We also note that these expressions take into account the frequency dependence of the effective LGR resistance \mbox{$R=R_0\sqrt{\omega/\omega_0}$}~\cite{Bobowski:2013, Bobowski:2015}.

\subsection{Impedance of a partially-filled LGR}
We now consider a LGR resonator of length $\ell$ with its bore containing an ESRR or a 1-D SRR array of length $z$.  The filling fraction of the loaded LGR is given by $\eta=z/\ell$.  Due to the geometry, the same net magnetic flux passes through the filled and unfilled regions of the LGR bore.  Therefore, by Faraday's law of induction, the emf induced across the empty and filled sections of the bore are equal which suggests that the LGR inductance can be modelled as two parallel inductances as follows
\begin{equation}
L=\frac{\mu_\mathrm{r}L_1L_2}{L_1+\mu_\mathrm{r}L_2},
\end{equation}
where $L_1=\mu_\mathrm{r}L_0/\eta$ is the inductance of the filled region and $L_2=L_0/(1-\eta)$ is the inductance of the empty region~\cite{Dubreuil:2017}.  This effective inductance can be re-expressed in the form \mbox{$L=L_0\left(\ell_1+j\ell_2\right)$} where
\begin{align}
\ell_1 &=\frac{\mu^\prime\left[\mu^\prime\left(1-\eta\right)+\eta\right]+\left(\mu^{\prime\prime}\right)^2\left(1-\eta\right)}{\left[\mu^\prime\left(1-\eta\right)+\eta\right]^2+\left[\mu^{\prime\prime}\left(1-\eta\right)\right]^2},\label{eq:ell1}\\
\ell_2 &=\frac{-\mu^{\prime\prime}\eta}{\left[\mu^\prime\left(1-\eta\right)+\eta\right]^2+\left[\mu^{\prime\prime}\left(1-\eta\right)\right]^2}.\label{eq:ell2}
\end{align}

Our goal is to calculate $Z_\mathrm{f}$ of the partially-filled LGR inductively coupled to $L_1$.  The equivalent circuit of Fig.~\ref{fig:mutual} remains valid if we make the replacements $L_0\to L$, $M_0\to M$, and $Z_\mathrm{e}\to Z_\mathrm{f}$.  The form of $L$ has already been determined, so we next consider the mutual inductance which is given by \mbox{$M=k\sqrt{L L_1}$} where $0\le k\le 1$ is a coupling constant determined by the position of the coupling loop relative to the bore of the LGR.  Using the above form of $L$, $M$ can be rewritten as \mbox{$M=M_0\sqrt{\ell_1+j\ell_2}$} where \mbox{$M_0=k\sqrt{L_0 L_1}$}.  By expressing $\sqrt{\ell_1+j\ell_2}$ as a complex exponential and making use of standard trigonometric identities, it is possible to write
\begin{equation}
\sqrt{\ell_1+j\ell_2}=\left\vert\ell\right\vert^{1/2}\left(\cos\frac{\phi}{2}+j\sin\frac{\phi}{2}\right),
\end{equation}
where
\begin{align}
\left\vert\ell\right\vert^2 &=\ell_1^2+\ell_2^2,\\
\cos\frac{\phi}{2} &=\sqrt{\frac{1+\ell_1/\left\vert\ell\right\vert}{2}},\label{eq:cos}\\
\sin\frac{\phi}{2} &=-\sqrt{\frac{1-\ell_1/\left\vert\ell\right\vert}{2}}.\label{eq:sin}
\end{align}
In Eqs.~(\ref{eq:cos}) and (\ref{eq:sin}), the positive root for \mbox{$\cos\left(\phi/2\right)$} has been chosen since $\ell_1>0$, while the negative root has been chosen for \mbox{$\sin\left(\phi/2\right)$} since $\ell_2<0$.  Combining the results above produces the following expression for the mutual inductance between the coupling loop and the SRR-loaded LGR
\begin{align}
M &=M_0\left(m_1+j m_2\right),\\
m_1 &=\left\vert\ell\right\vert^{1/2}\cos\frac{\phi}{2},\\
m_2 &=\left\vert\ell\right\vert^{1/2}\sin\frac{\phi}{2}.
\end{align}

Returning to the equivalent circuit of Fig.~\ref{fig:mutual}, detailed analysis of \mbox{$Z_\mathrm{f}=R_\mathrm{f}+j X_\mathrm{f}$} leads to
\begin{widetext}
\begin{align}
R_\mathrm{f}&=-\frac{\left(\omega M_0\right)^2}{R_0}\frac{\left[\left(m_2^2-m_1^2\right)\left(\sqrt{\dfrac{\omega}{\omega_0}}-Q_0\dfrac{\omega}{\omega_0}\ell_2\right)-2m_1m_2Q_0\left(\dfrac{\omega}{\omega_0}\ell_1-\dfrac{\omega_0}{\omega}\right)\right]}{\left[\sqrt{\dfrac{\omega}{\omega_0}}-Q_0\dfrac{\omega}{\omega_0}\ell_2\right]^2+Q_0^2\left[\dfrac{\omega}{\omega_0}\ell_1-\dfrac{\omega_0}{\omega}\right]^2},\label{eq:Rf}\\
X_\mathrm{f}&=\omega L_1+\frac{\left(\omega M_0\right)^2}{R_0}\frac{\left[2m_1m_2\left(\sqrt{\dfrac{\omega}{\omega_0}}-Q_0\dfrac{\omega}{\omega_0}\ell_2\right)+\left(m_2^2-m_1^2\right)Q_0\left(\dfrac{\omega}{\omega_0}\ell_1-\dfrac{\omega_0}{\omega}\right)\right]}{\left[\sqrt{\dfrac{\omega}{\omega_0}}-Q_0\dfrac{\omega}{\omega_0}\ell_2\right]^2+Q_0^2\left[\dfrac{\omega}{\omega_0}\ell_1-\dfrac{\omega_0}{\omega}\right]^2}.\label{eq:Xf}
\end{align}
\end{widetext}
Note that, if either $\mu^\prime=1$ and $\mu^{\prime\prime}=0$ or $\eta=0$, then \mbox{$\ell_1=\cos\left(\phi/2\right)=m_1=1$} and \mbox{$\ell_2=\sin\left(\phi/2\right)=m_2=0$}.  It follows that, in both cases, $R_\mathrm{f}$ reduces to $R_\mathrm{e}$ and $X_\mathrm{f}$ reduces to $X_\mathrm{e}$, as expected. 

In the above analysis, the modified LGR inductance due to the presence of the ESRR/SRR array also affected the mutual inductance between the coupling loop and the LGR.  This complication can be avoided if one chooses to instead capacitively couple to the LGR.  Capacitive coupling methods are described and analyzed in Ref.~[\onlinecite{Rinard:1993}].  

\subsection{VNA reflection coefficient}
In an experiment, one typically uses a vector network analyzer (VNA) to measure the reflection coefficient \mbox{$S_{11,\mathrm{e/f}}=\left(Z_\mathrm{e/f}-Z_0\right)/\left(Z_\mathrm{e/f}+Z_0\right)$} where $Z_0=\SI{50}{\ohm}$ is the characteristic impedance of the transmission line.  This work will focus primarily on the magnitude of the $S_{11,\mathrm{e/f}}$ which can be expressed as
\begin{equation}
\left\vert S_{11,\mathrm{e/f}}\right\vert=\frac{\sqrt{\left(\dfrac{\left\vert Z_\mathrm{e/f}\right\vert^2}{Z_0^2}-1\right)^2+\left(2\dfrac{X_\mathrm{e/f}}{Z_0}\right)^2}}{\left(\dfrac{\left\vert Z_\mathrm{e/f}\right\vert^2}{Z_0^2}+1\right)+2\dfrac{R_\mathrm{e/f}}{Z_0}},\label{eq:S11mag}
\end{equation}
where \mbox{$\left\vert Z_\mathrm{e/f}\right\vert^2=R_\mathrm{e/f}^2+X_\mathrm{e/f}^2$}.  

If $L_1$ for a particular coupling loop is known, then a measurement of $\left\vert S_{11,\mathrm{e}}\right\vert$ can be used to determine $\omega_0$, $Q_0$, and $M_0^2/R_0$ of the empty LGR.  A follow-up measurement of $\left\vert S_{11,\mathrm{f}}\right\vert$ for the SRR-loaded LGR then depends only on $\eta$, $\omega_\mathrm{S}$, $\omega_\mathrm{P}$, and $\gamma_\mathrm{S}$.  This pair of measurements, therefore, allows one to experimentally determine all of the quantities that parameterize the complex permeability of the ESRR/SRR array filling the bore of the LGR.

\begin{figure*}
\begin{tabular}{lr}
(a)\includegraphics[width=7.7 cm]{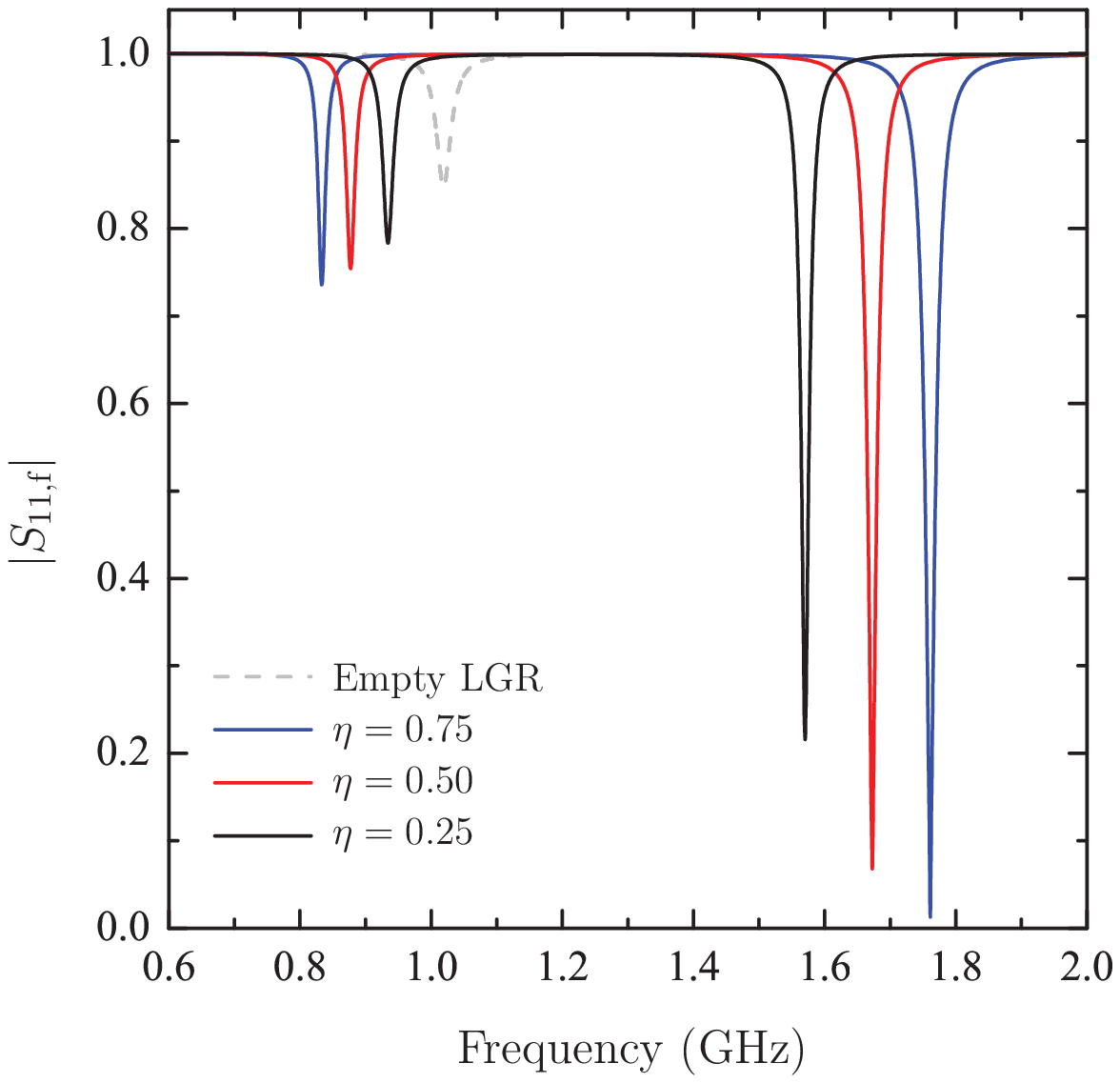}\quad~ & ~\quad(b)\includegraphics[width=7.7 cm]{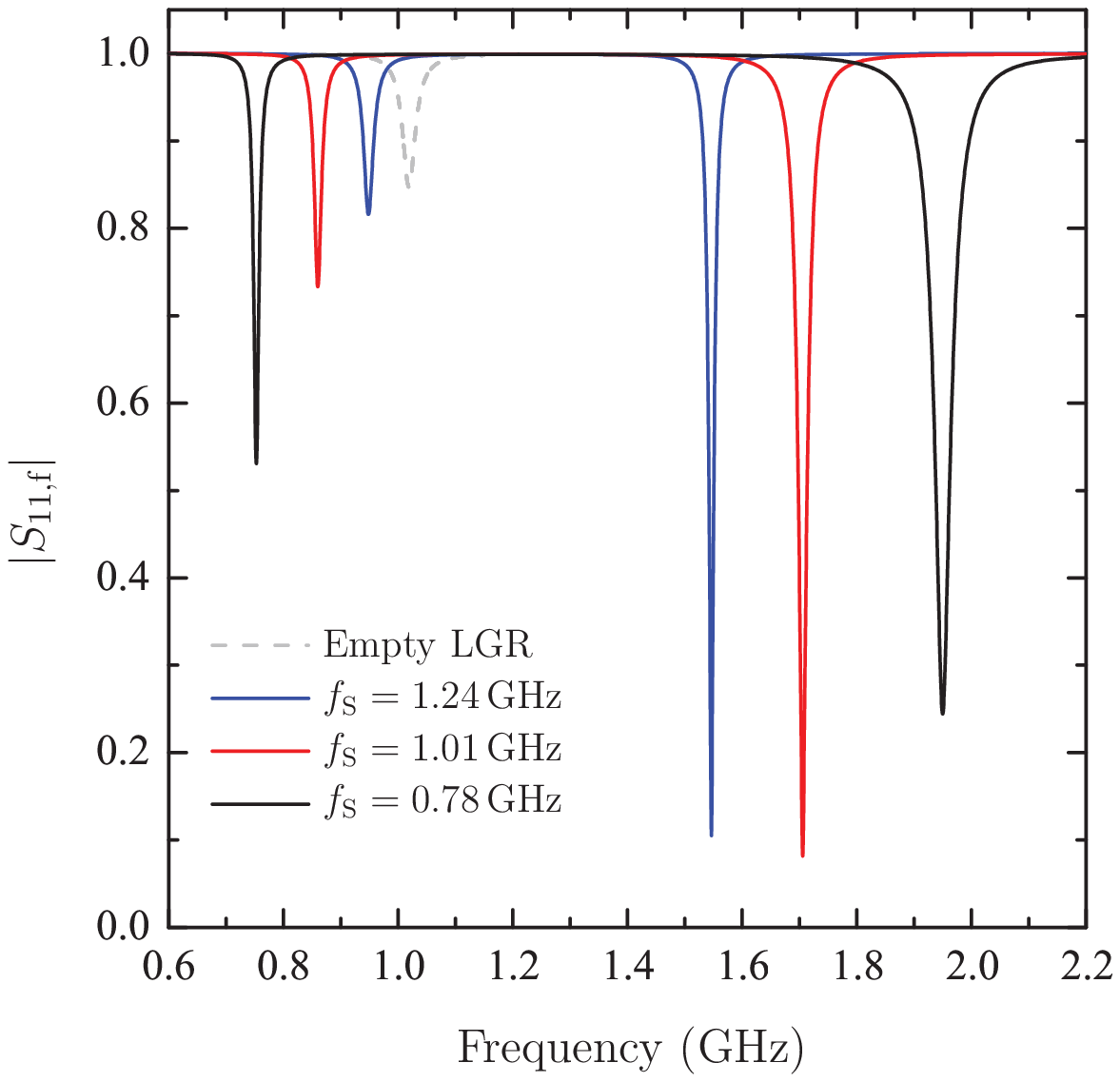}\\
~ & ~\\
(c)\includegraphics[width=7.7 cm]{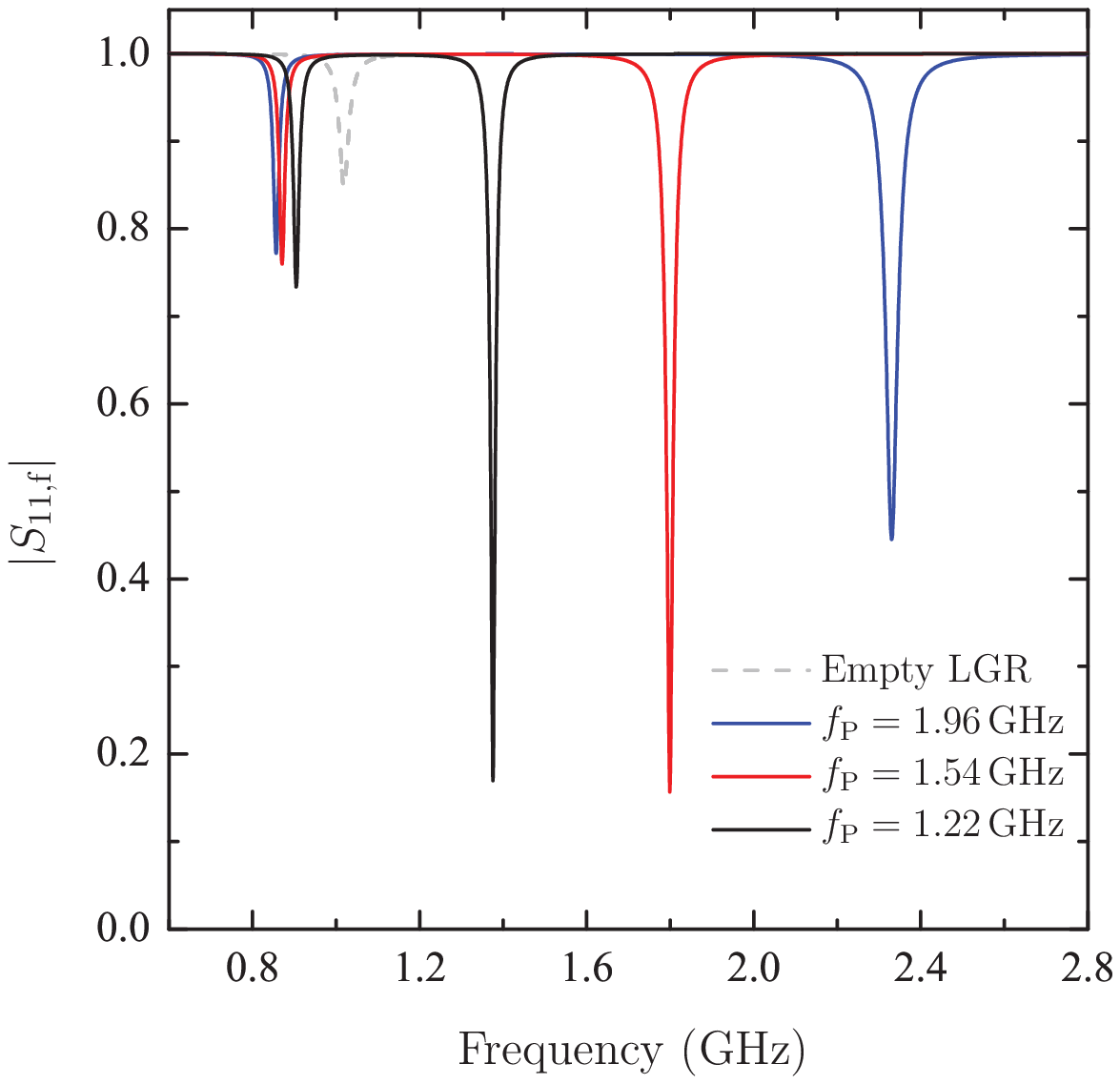}\quad~ & ~\quad(d)\includegraphics[width=7.7 cm]{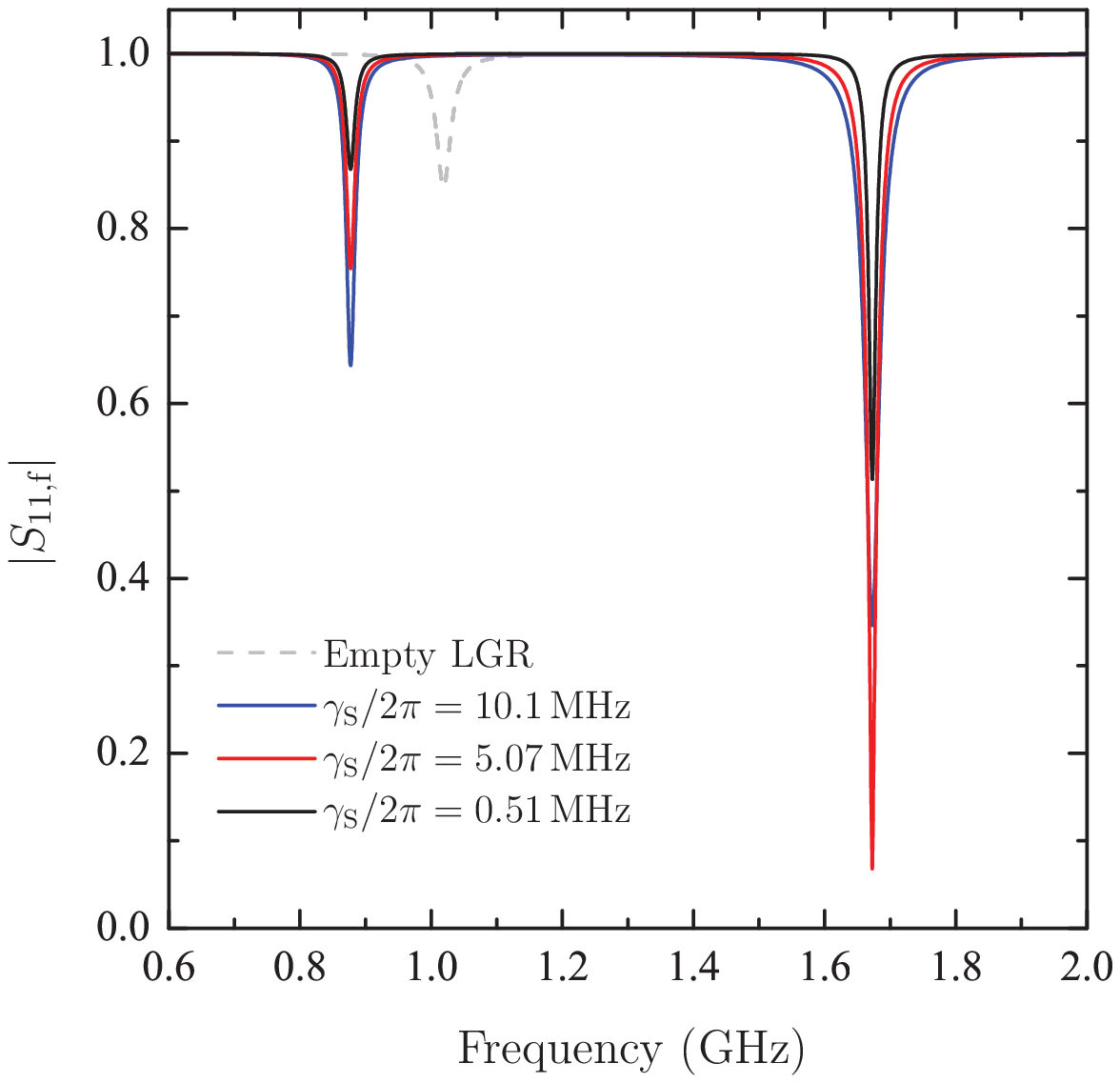}
\end{tabular}
\caption{\label{fig:scans}Plots of $\left\vert S_{11,\mathrm{f}}\right\vert$ versus frequency.  In all four plots, values of $L_1=\SI{12}{\nano\henry}$ and $M_0^2/R_0 = \SI{0.05}{\nano\henry^2/\milli\ohm}$ were used.  The dashed line represents $\left\vert S_{11,\mathrm{e}}\right\vert$ of an empty LGR with $f_0=\SI{1}{\giga\hertz}$ and $Q_0=500$. (a) The calculated response for various filling factors $\eta$ when $f_\mathrm{S}=\SI{1.05}{\giga\hertz}$, $f_\mathrm{P}=\SI{1.44}{\giga\hertz}$, and $\gamma_\mathrm{S}/2\pi=\SI{5.07}{\mega\hertz}$. (b) The calculated response for various resonant frequencies $f_\mathrm{S}$ when $\eta=0.50$, $f_\mathrm{P}=\SI{1.44}{\giga\hertz}$, and $\gamma_\mathrm{S}/2\pi=\SI{5.07}{\mega\hertz}$. (c) The calculated response for various plasma frequencies $f_\mathrm{P}$ when $\eta=0.50$, $f_\mathrm{S}=\SI{1.05}{\giga\hertz}$, and $\gamma_\mathrm{S}/2\pi=\SI{5.07}{\mega\hertz}$.  (d) The calculated response for various damping constants $\gamma_\mathrm{S}/2\pi$ when $\eta=0.50$, $f_\mathrm{S}=\SI{1.05}{\giga\hertz}$, and $f_\mathrm{P}=\SI{1.44}{\giga\hertz}$.}
\end{figure*}

Figure~\ref{fig:scans} shows the calculated $\left\vert S_{11,\mathrm{f}}\right\vert$ as a function of frequency as (a) $\eta$ and the permeability parameters (b) \mbox{$f_\mathrm{S}=\omega_\mathrm{S}/2\pi$}, (c) \mbox{$f_\mathrm{P}=\omega_\mathrm{P}/2\pi$}, and (d) $\gamma_\mathrm{S}/2\pi$ are scanned.  
In these plots, \mbox{$f_0=\omega_0/2\pi=\SI{1.0}{\giga\hertz}$} and $Q_0=500$ have been assumed for the LGR.  For the coupling loop and mutual inductance, we have assumed \mbox{$L_1=\SI{12}{\nano\henry}$} and \mbox{$M_0^2/R_0 = \SI{0.05}{\nano\henry^2/\milli\ohm}$}.  Except for $Q_0$, all of these values correspond to values that were measured in the experiments presented in Sec.~\ref{sec:expt}.  A reduced value of the LGR quality factor $Q_0$ was used so as to broaden the resonances in Fig.~\ref{fig:scans} for clarity.  The dashed line in the figures represents the resonance due to an empty LGR.  The nominal ESRR/SRR array parameters used in the plots were $\eta=0.50$, $f_\mathrm{S}=\SI{1.05}{\giga\hertz}$, $f_\mathrm{P}=\SI{1.44}{\giga\hertz}$, and $\gamma_\mathrm{S}/2\pi=\SI{5.07}{\mega\hertz}$.  

The figures show that the loaded LGR produces a double resonance.  As expected, as the filling factor $\eta$ in Fig.~\ref{fig:scans}(a) is increased, the separation between the two resonances increases.  The positions of the low- and high-frequency resonances are both affected by $\eta$ and $f_\mathrm{S}$.  However, because the relative heights of the two resonances have distinct dependencies on $\eta$ and $f_\mathrm{S}$, fits to $\left\vert S_{11,\mathrm{f}}\right\vert$ measurements can be used to reliably extract both parameters.  Figure~\ref{fig:scans}(b) shows that this method does not require $f_0$ of the LGR to be well matched to $f_\mathrm{S}$ of the ESRR/SRR array.  Clear low- and high-frequency resonances exist even when $f_\mathrm{S}$ is more than $20$\% below or above $f_0$.  The plasma frequency $f_\mathrm{P}$ has very little effect on the low-frequency resonance, but plays a major role in setting the position of the high-frequency resonance.  The damping constant $\gamma_\mathrm{S}$ affects only the widths and relative heights of the two resonances.

Finally, we note that the values of $L_1$ and $M_0^2/R_0$ used correspond to a LGR that is overcoupled.  As a result, for the empty LGR, \mbox{$\left\vert S_{11,\mathrm{e}}\right\vert > 0.8$} at the resonant frequency.  The coupling can be tuned by moving the position of the coupling loop relative to the bore of the LGR.  At a ``critical'' coupling, $\left\vert S_{11,\mathrm{e}}\right\vert$ approaches zero as $f$ approaches  the resonance.  As discussed in detail in Sec.~\ref{sec:expt}, the coupling loop position was set so as to produced the strongest resonances when the LGR was loaded with an ESRR/SRR array.  With the coupling set in this way, the LGR becomes overcoupled once its bore has been emptied.

\section{Charactersizing a 3-D SRR Array}\label{sec:multiple}
Section~\ref{sec:SRRLGR} considered a single ESRR or a 1-D SRR array loaded in the bore of a LGR.  Since the resonant properties of the LGR are approximately independent of $\ell$, it can be made arbitrarily long without seriously affecting either $\omega_0$ or $Q_0$~\footnote{The resonant frequency $\omega_0$ will decrease slightly as the LGR length $\ell$ is increased}.  Therefore, it is possible to use the methods described in the preceding section to study 1-D arrays of SRRs with any number of unit cells stacked along the length of the LGR axis.

Studying 3-D arrays of SRRs is, however, more challenging.  For example, loading a LGR with a $2\times 2$ configuration of SRRs in the cross-section of the bore, requires the bore area to quadruple (see Table~\ref{tab:LGRdesigns}).       
\begin{table}
\caption{\label{tab:LGRdesigns}LGR designs for characterizing a 1-D SRR array and various 3-D arrays of SRRs.  The drawings are not to scale.  All of the designs use a SRR unit cell size of \mbox{$a=\SI{12.70}{\milli\meter}$} and result in a LGR resonant frequency of \mbox{$f_0=\SI{1.0}{\giga\hertz}$}.}
\begin{ruledtabular}
\begin{tabular}{rl}
\includegraphics[width=2.35 cm]{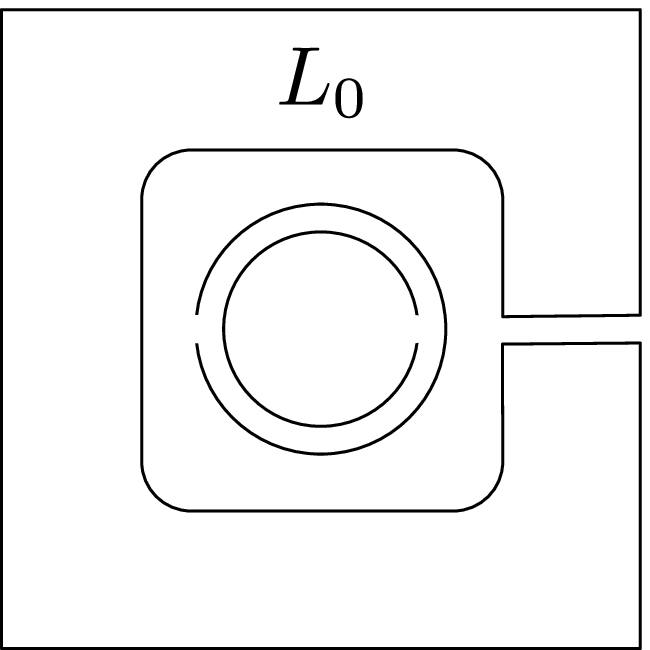} & \begin{tabular}[b]{@{}l@{}}$\displaystyle{f_0=\frac{1}{2\pi}\frac{c}{a}\sqrt{\frac{t}{w}}}$\\~\\$x=a$\\$t=\SI{0.76}{\milli\meter}$\\$w=\SI{10.77}{\milli\meter}$\\[-8 pt]~\end{tabular}\\
\includegraphics[width=4.2 cm]{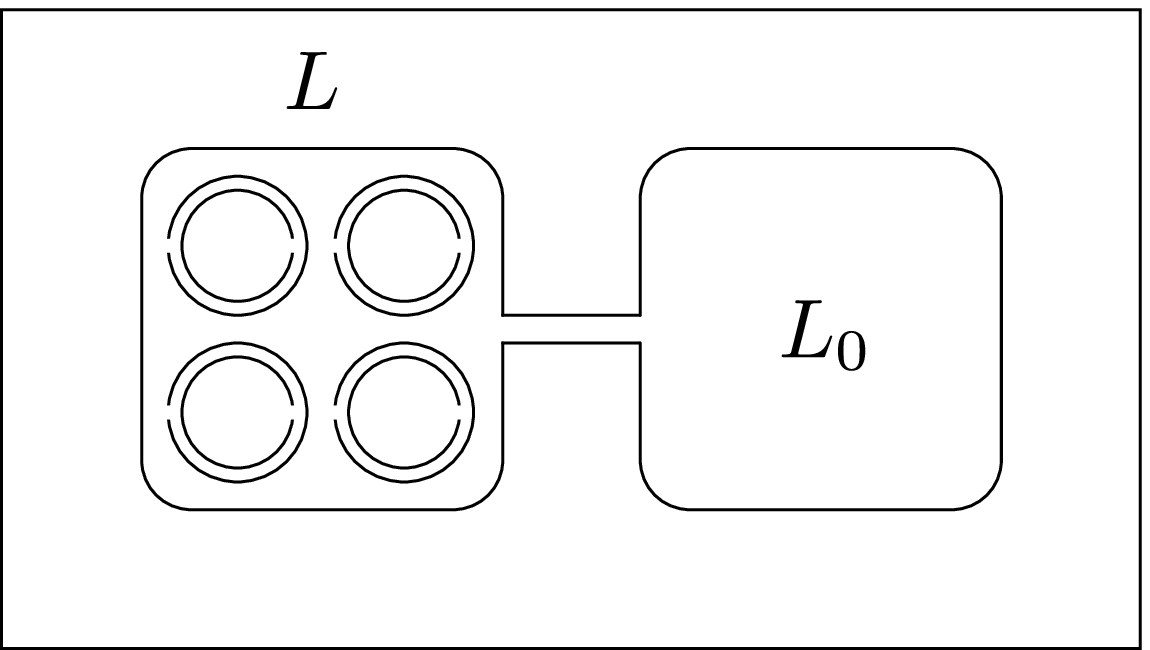} & \begin{tabular}[b]{@{}l@{}}$\displaystyle{f_0=\frac{1}{2\pi}\frac{c}{a}\sqrt{\frac{t}{2w}}}$\\~\\$x=2a$\\$t=\SI{1.52}{\milli\meter}$\\$w=\SI{10.80}{\milli\meter}$\\[-8 pt]~\end{tabular}\\
\includegraphics[width=6 cm]{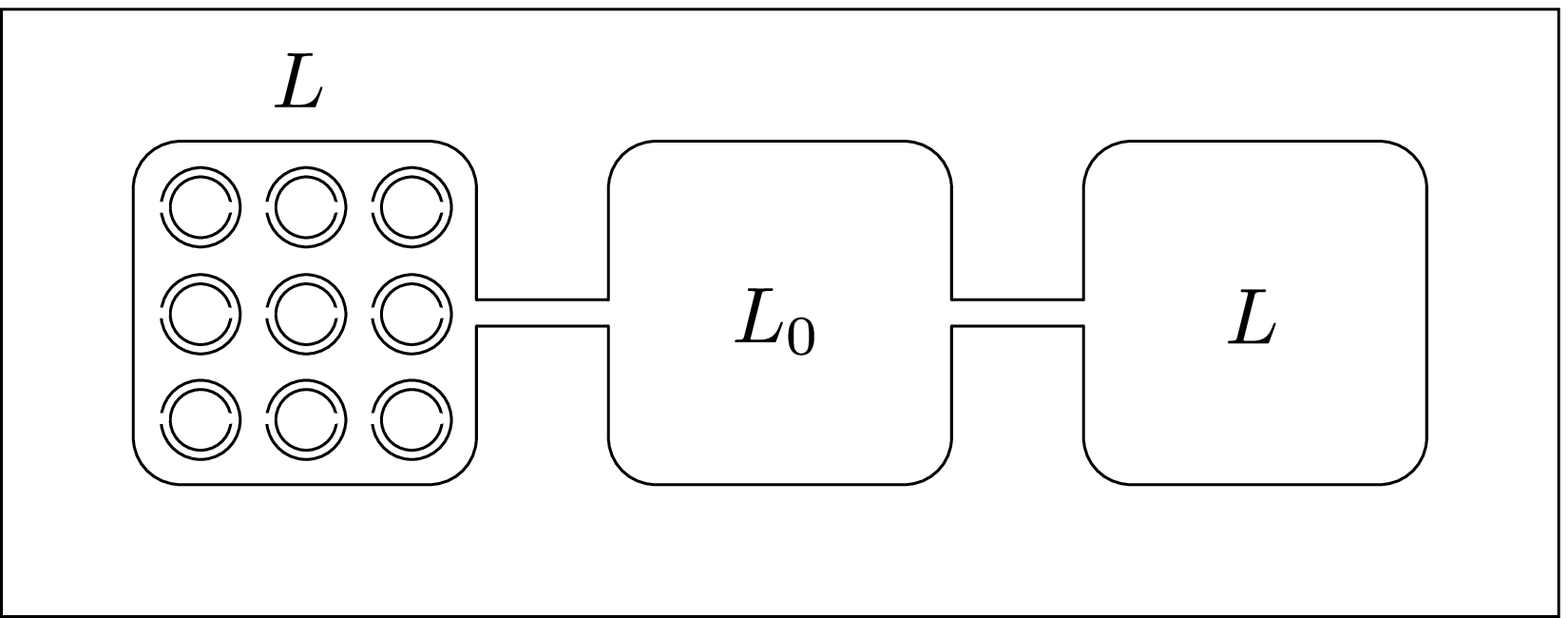} & \begin{tabular}[b]{@{}l@{}}$\displaystyle{f_0=\frac{1}{2\pi}\frac{c}{a}\sqrt{\frac{t}{3w}}}$\\~\\$x=3a$\\$t=\SI{2.03}{\milli\meter}$\\$w=\SI{9.60}{\milli\meter}$\\[-8 pt]~\end{tabular}\\
\includegraphics[width=6 cm]{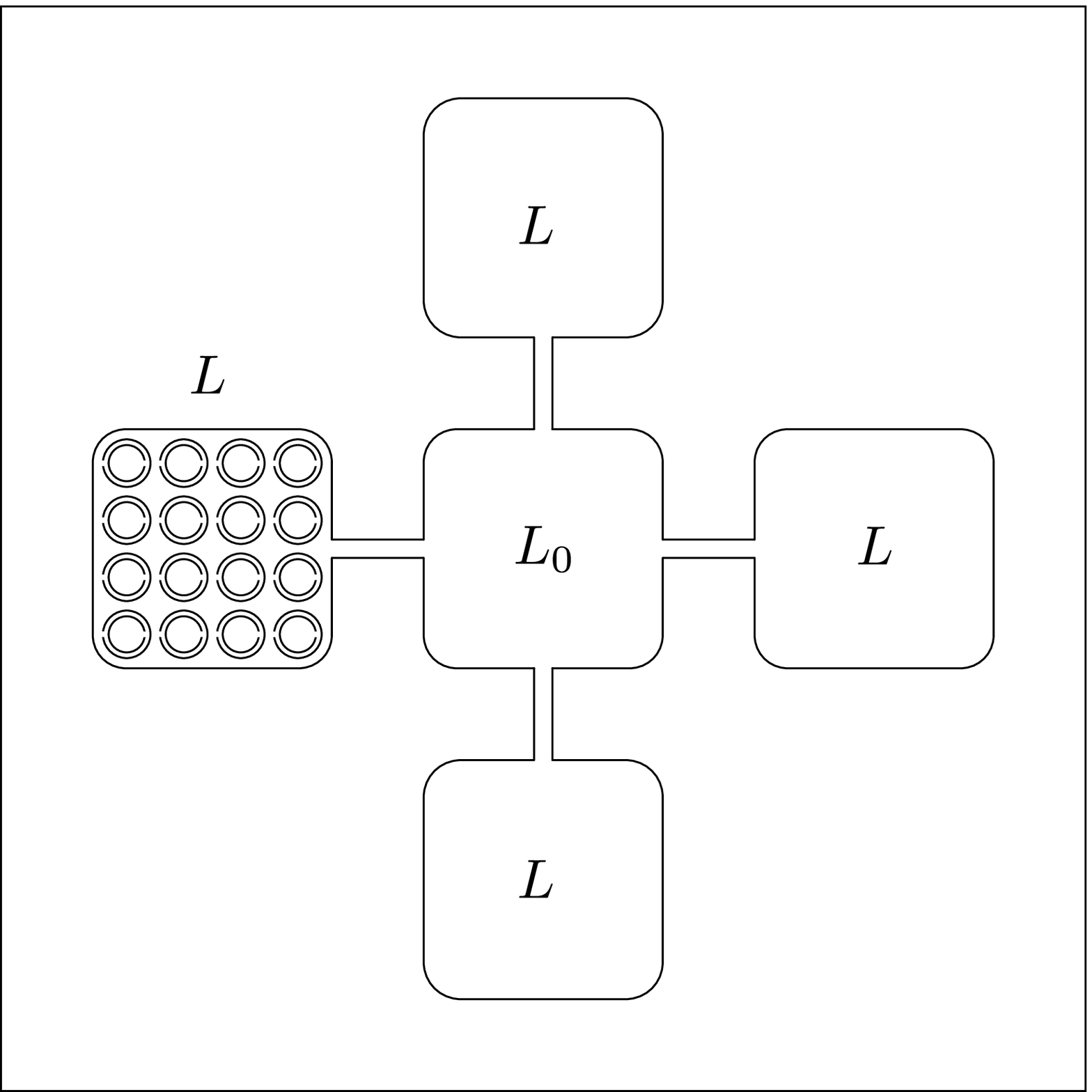} & \begin{tabular}[b]{@{}l@{}}$\displaystyle{f_0=\frac{1}{2\pi}\frac{c}{a}\sqrt{\frac{5t}{16w}}}$\\~\\$x=4a$\\$t=\SI{2.54}{\milli\meter}$\\$w=\SI{11.20}{\milli\meter}$\\[-8 pt]~\end{tabular}
\end{tabular}
\end{ruledtabular}
\end{table}
As Eq.~(\ref{eq:w0}) shows, a doubling of the bore dimension $x$ will cause $\omega_0$ of the LGR to decrease by a factor of two.  A way of increasing the cross-sectional area of the bore while simultaneously  maintaining the same resonant frequency is needed.  Fortunately, Hyde, Froncisz, and co-workers have described multi-loop, multi-gap LGR structures that can be used to satisfy this design requirement~\cite{Wood:1984, Froncisz:1986, Hyde:1989a, Rinard:1994, Sidabras:2007, Hyde:2007}.

\subsection{Multi-loop, multi-gap LGRs}
We start by considering the two-loop, one-gap LGR shown in the second row of Table~\ref{tab:LGRdesigns}~\cite{Froncisz:1986}.  If it is assumed that there is equal magnetic flux in each of the loops, then the equivalent circuit of the structure near resonance is a parallel combination of the inductance of each loop in series with the gap capacitance.  
%In general, the inductances of the left and right loops can be different.  
The right-hand column of Table~\ref{tab:LGRdesigns} gives an approximate expression for $f_0$ of the LGR when the two loops are assumed to be identical.  In this expression, $a$ is the size of the SRR unit cell.  For the two-loop, one-gap LGR, the bore size is $x=2a$ such that a $2\times 2\times N$ array of SRRs can be loaded into the LGR.  Here, $N$ is the number of SRRs stacked along the length of the LGR bore.  The table also gives values for the gap height $t$ and gap width $w$ that would result in \mbox{$f_0\approx\SI{1}{\giga\hertz}$}.  For all of the LGR designs considered in Table~\ref{tab:LGRdesigns}, we assume a SRR unit cell size of \mbox{$a=\SI{12.70}{\milli\meter}$}.  Although the diagrams in the table suggest the SRRs are getting smaller as the array size is expanded, we are really imagining that the SRR size remains constant and the cross-sectional area of the LGR bore is increasing.

Increasing the LGR bore size further to accommodate a $3\times 3\times N$ array of SRRs can be done using a three-loop, two-gap LGR~\cite{Wood:1984, Hyde:1989a}.  Wood {\it et al.\@} report that this structure supports two resonant modes.  In the low-frequency (fundamental) mode, there is no magnetic flux in the central loop and magnetic field lines form closed paths by passing through each of the outer loops.  It is, however, the high-frequency mode that is of interest here.  In this mode, no magnetic flux is shared between the two outer loops.  Rather, magnetic field lines form closed paths by passing through the central loop and one of the outer loops.  Again, Table~\ref{tab:LGRdesigns} gives an expression for $f_0$ valid when the three loops are identical with $x=3a$.  The dimensions given would produce a LGR with \mbox{$f_0\approx\SI{1}{\giga\hertz}$}.

Finally, the last row in the table shows a five-loop, four-gap LGR~\cite{Rinard:1994, Sidabras:2007, Hyde:2007}.  Again, in the mode of interest, no magnetic flux is shared amongst any of the outside loops.  Each magnetic field line passes through the central loop and only one of the four outer loops.  If the five loops are identical with $x=4a$, Table~\ref{tab:LGRdesigns} gives an approximate expression for $f_0$ and LGR dimensions that would set \mbox{$f_0\approx\SI{1}{\giga\hertz}$}.  In proposed design, $x=\SI{5.1}{\centi\meter}$ and the free-space wavelength at $f_0$ is \SI{30}{\centi\meter}.  Increasing the bore size of the LGR any further, while maintaining a \SI{1}{\giga\hertz} resonant frequency, would likely result in a breakdown of the lumped-element model of the LGR.   

Using any of the multi-loop, multi-gap LGRs, it would be possible to inductively couple to any one of the loops or capacitively couple to any one of the gaps.  Once the geometry of the coupling is set, an equivalent circuit of the type shown in Fig.~\ref{fig:mutual} would need to be developed which would allow both $Z_\mathrm{e}$ and $Z_\mathrm{f}$ to be calculated.  Using Eq.~(\ref{eq:S11mag}), these impedances determine the forms of $\left\vert S_{11,\mathrm{e}}\right\vert$ and $\left\vert S_{11,\mathrm{f}}\right\vert$ which can be fit to experimental data to extract the effective permeability of the SRR array.

\subsection{Multi-loop, multi-gap LGR circuit model}
\begin{figure}
\includegraphics[width=8 cm]{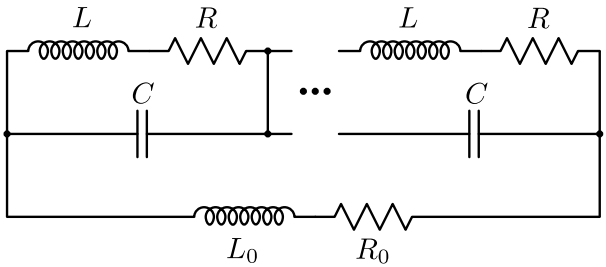}% Here is how to import EPS art
\caption{\label{fig:multiloop}Equivalent circuit used to model $n$-loop, $m$-gap LGRs.  This model assumes that $n-1$ identical outer loops of inductance $L$ share all of their magnetic flux  with a single central loop of inductance $L_0$.  For each of the outer loops, there is a corresponding $L$, $R$, and $C$ Kirchhoff loop in the equivalent circuit model.}
\end{figure}

We now give a general lumped-element circuit model for an $n$-loop, $m$-gap LGR~\cite{Wood:1984}.  In all but the one-loop, one-gap LGR, the magnet flux in each of the loops labelled $L$ (referred to as ``outer'' loops) is shared with a single loop labelled $L_0$ (the ``central'' loop).  As a result, the emf associated with the central loop must the equal to the sum of the emfs associated with outer loops.  In multi-gap LGRs, the current circulating the central loop must pass through a series combination of the gaps.  These observations suggest that the circuit model shown in Fig.~\ref{fig:multiloop} is valid near the resonance of interest.  
In the circuit, the number of \mbox{$\left(j\omega L+R\right)\mathbin{\parallel}1/j\omega C$} branches is equal to $n-1$, i.e.\@ the number of outer loops.

\section{Experimental Demonstration using a Toroidal LGR}\label{sec:expt}

\subsection{Apparatus}
\begin{figure}
\includegraphics[width=7.7 cm]{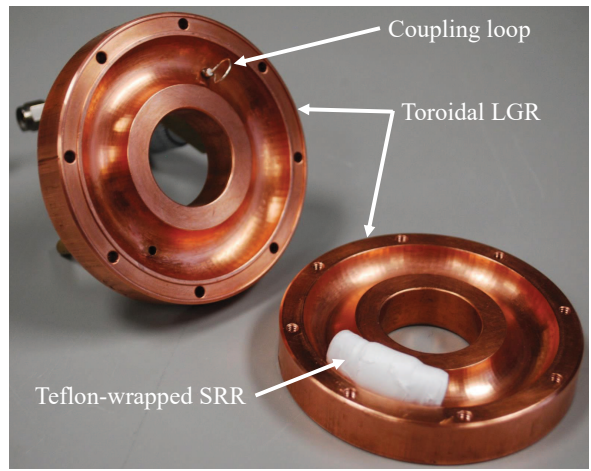}% Here is how to import EPS art
\caption{\label{fig:SRRinLGR}Digital photograph of the disassembled TLGR with its bore loaded with an ESRR that has been wrapped in Teflon tape.  Also visible is the coupling loop suspended within the TLGR bore.  For scale, the outer diameter of the TLGR is \SI{7.6}{\centi\meter} $\left(\SI{3}{in}.\right)$.}
\end{figure}

This section describes the first demonstration of our proposed method used to determine the complex permeability of a single ESRR.  For these measurements, we used a toroidal LGR (TLGR) that is shown in Fig.~\ref{fig:SRRinLGR}.  
As described elsewhere, the main advantage of the TLGR is that the magnetic flux is strongly confined within the bore of the resonator such that no additional EM shielding is required to preserve a high quality factor~\cite{Bobowski:2016}.  That the bore axis is circular and the magnetic field strength is not uniform across the cross-section of the bore are the main disadvantages.  The resonant frequency and unloaded $Q$ of the TLGR used are $f_0\approx \SI{1.0}{\giga\hertz}$ and $Q_0\approx 2400$, respectively~\cite{Bobowski:2016}. 

Also shown in Fig.~\ref{fig:SRRinLGR}, is the coupling loop made from UT-085 semi-rigid coaxial cable.  After stripping away a length of the outer conductor and dielectric, the coax was passed through a small hole that provides access to the bore of the resonator.  The center conductor was then bent into a loop and its end was soldered to the outer conductor of the coaxial cable.  The diameter of the TLGR bore is \SI{13}{\milli\meter} $\left(\SI{0.5}{in}.\right)$ and the coupling loop diameter is approximately \SI{7}{\milli\meter}.  The coupling strength was adjusted by rotating the coupling loop with respect to the axis of the TLGR bore.  Using this method, the coupling strength could be tuned to achieve anything between a strongly undercoupled to a strongly overcoupled resonator. 

For our measurements, we constructed a crude ESRR of the type described in Fig.~3 of Ref.~[\onlinecite{Pendry:1999}] and shown in cross-section in Fig.~\ref{fig:1loop1gap} of this paper.  First, a length of polyethylene tubing was wrapped around an aluminum cylinder and heated using a heat gun.  This process was repeated until we had a short section of tubing with a curvature that matched that of the bore of the TLGR.  Once the heating process was completed, the curved section of tubing had an elliptical cross-section with major- and minor-axes of \SI{11.0}{\milli\meter} and \SI{8.8}{\milli\meter}, respectively.  Next, this section of tubing was completely wrapped with \SI[number-unit-product=\text{-}]{36}{\micro\meter} thick copper tape.  A razor was then used to cut a slit in the copper foil along the inside radius of the curved tube.  To produce an approximately uniform spacing $d$ between the concentric cylinders of the ESRR, the entire assembling was then wrapped with several layers of Teflon tape.  Next, an outer layer of copper tape was added on top of the Teflon. A slit was cut into this second layer of copper along the outer radius of the curved ESRR.  Orienting the slits in this way minimizes the interaction between the ESRR and the fringing electric fields from the gap of the TLGR.  Finally, the ESRR was wrapped with several more layers of Teflon tape to electrically isolate it from the bore of the TLGR.  To position the ESRR in the center of the bore of the TLGR, a thick layer of Teflon tape was added at its midpoint.

\begin{figure}
\includegraphics[width=7.7 cm]{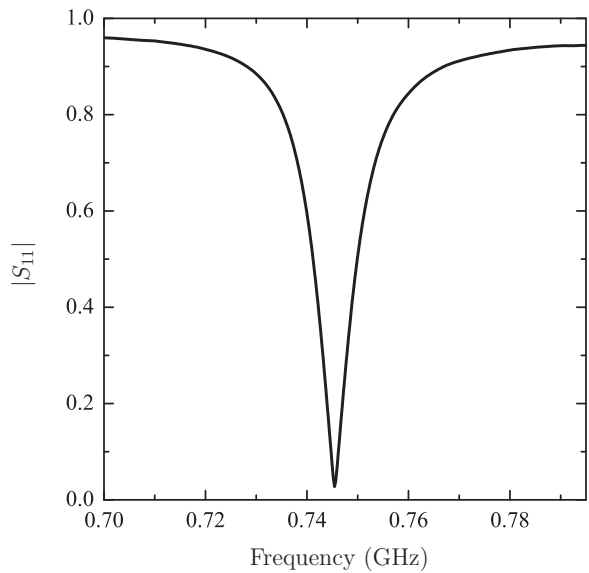}% Here is how to import EPS art
\caption{\label{fig:unshieldedSRR}Resonance of the ESRR when suspended in air.  The coupling loop was positioned so as to achieve near-critical coupling at the resonant frequency which was measured to be \SI{745.4}{\mega\hertz}.}
\end{figure}

Several ESRRs were made in the way described above until one with a suitable resonance frequency $f_\mathrm{S}$ was found.  With the ESRR suspended in air, a coupling loop was positioned so as to achieve near-critical coupling.  A VNA was used to measure the reflection coefficient of the inductively-coupled ESRR and the result is shown in Fig.~\ref{fig:unshieldedSRR}. 
The resonant frequency of the unshielded ESRR that was used for the remainder of our measurements was determined to be \SI{745.4}{\mega\hertz}.
\begin{comment}
Because this measurement was made without any EM shielding surrounding the SRR, radiative losses make the width of the resonance larger than the expected value given by the damping constant $\gamma_\mathrm{S}/2\pi$.
\end{comment}

The measurement described above, and all other reported measurements, were made using an Agilent E5061A \SI{300}{\kilo\hertz} to \SI{1.5}{\giga\hertz} VNA.  The VNA was calibrated using the Agilent 85033E calibration kit.  All reported measurements were made with the VNA output power set to \SI{10}{dBm}.

\subsection{Coupling loop inductance}\label{sec:loop}    
The reactances $X_\mathrm{e}$ and $X_\mathrm{f}$ of the empty and partially-filled LGR given by Eqs.~(\ref{eq:Xe}) and (\ref{eq:Xf}) depend on the inductance $L_1$ of the coupling loop.  This section describes the method used to determine $L_1$ of our coupling loop.

The impedance of a lossless transmission line of length $s$ terminated by an inductance $L_1$ is
\begin{equation}
Z_\mathrm{in}=jZ_0\frac{\omega L_1/Z_0+\tan qf}{1-\left(\omega L_1/Z_0\right)\tan qf},
\end{equation} 
where \mbox{$q\equiv 2\pi s\sqrt{\varepsilon_\mathrm{r}}/c$} and $\varepsilon_\mathrm{r}$ is the dielectric constant of the insulator filling the coaxial cable.  This expression can be used to determine the real and imaginary components of \mbox{$S_{11}=\left(Z_\mathrm{in}-Z_0\right)/\left(Z_\mathrm{in}+Z_0\right)$}.

\begin{figure}
\includegraphics[width=7.7 cm]{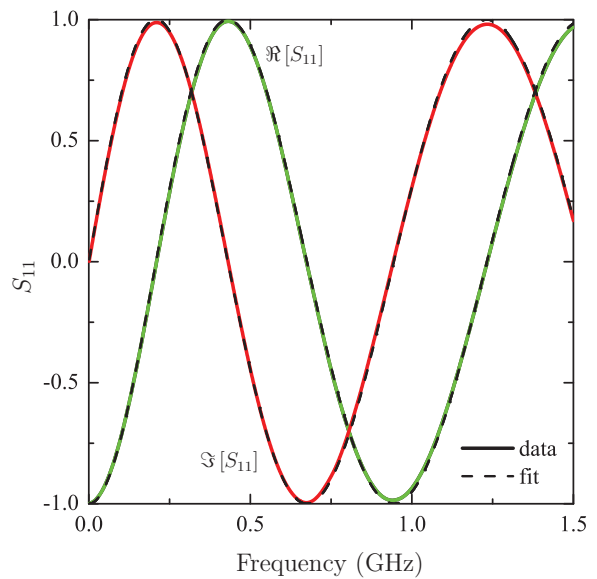}% Here is how to import EPS art
\caption{\label{fig:L1}The real (green) and imaginary (red) components of the $S_{11}$ reflection coefficient of a short length of coaxial transmission line terminated by a coupling loop of inductance $L_1$.  The dashed lines are simultaneous fits to the data.}
\end{figure}

Our measurements of $\Re\left[S_\mathrm{11}\right]$ and $\Im\left[S_\mathrm{11}\right]$ of a short section of transmission line terminated by a coupling loop are shown in Fig.~\ref{fig:L1}.
Simultaneous nonlinear fits to the data yielded the best-fit parameters \mbox{$L_1=\SI{12.00\pm 0.02}{\nano\henry}$} and \mbox{$q=\SI{2.3131\pm 0.0007}{\nano\second}$}.  This value of $L_1$ is used for the remainder of the analysis.

\subsection{Empty and ESRR-loaded toroidal LGR}
To determine the permeability of the ESRR, it was necessary to measure both $\left\vert S_{11,\mathrm{e}}\right\vert$ of the empty TLGR and $\left\vert S_{11,\mathrm{f}}\right\vert$ of the ESRR-loaded TLGR.  First, the ESRR was placed into the bore of the TLGR and the two halves of the resonator were assembled.  The ESRR was positioned as far from the coupling loop as possible.  The coupling loop orientation was then adjusted while observing $\left\vert S_{11,\mathrm{f}}\right\vert$ on the VNA display.  The maximum signal was achieved when the plane of the coupling loop was perpendicular to the axis of the TLGR bore.  This is the orientation, which corresponds to maximum coupling, that is shown in Fig.~\ref{fig:SRRinLGR}.  Using a set screw, the coupling loop was then fixed in this position for the remainder of the the experiment and a measurement of $\left\vert S_{11,\mathrm{f}}\right\vert$ was recorded.  Next, the two halves of the TLGR were separated, the ESRR was removed, and the resonator was reassembled with an empty bore.  Without making any changes to the coupling loop orientation, a measurement of $\left\vert S_{11,\mathrm{e}}\right\vert$ was recorded.

\begin{figure}
\includegraphics[width=7.7 cm]{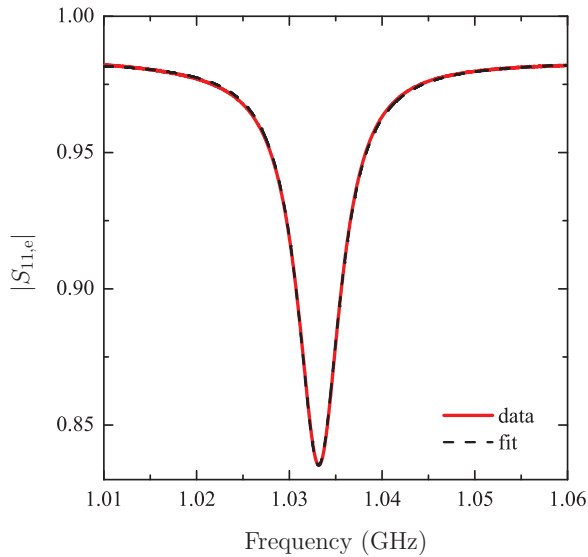}
\caption{\label{fig:empty}Magnitude of the reflection coefficient from an inductively coupled TLGR with an empty bore.  The coupling loop was oriented to achieve the maximum possible mutual inductance which results in an overcoupled resonator.  The dashed line is a fit to the data.}
\end{figure}

The data were analyzed in the opposite order that they were collected.  First, the $\left\vert S_{11,\mathrm{e}}\right\vert$ data were fit to Eq.~(\ref{eq:S11mag}) while using Eqs.~(\ref{eq:Re}) and (\ref{eq:Xe}) for the real and imaginary components of the impedance $Z_\mathrm{e}$ of the empty TLGR.  For these fits, the value given above for $L_1$ and \mbox{$Z_0=\SI{50}{\ohm}$} were used.  To empirically account for losses (ohmic and dielectric) along the length of the transmission line used to construct the coupling loop, an additional parameter $b$ was subtracted from $\left\vert S_{11,\mathrm{e}}\right\vert$ before executing the nonlinear fit routine.  As shown in Fig.~\ref{fig:empty}, the fit to the data is excellent.
The extracted parameters were \mbox{$f_0=\SI{1.029017\pm 0.000003}{\giga\hertz}$}, \mbox{$Q_0=\SI{2425\pm 1}{}$}, \mbox{$M_0^2/R_0=\SI{0.05076\pm 0.00002}{\nano\henry^2\per\milli\ohm}$}, and \mbox{$b=\SI{0.01627\pm 0.00002}{}$}.  The results for $f_0$ and $Q_0$ are consistent with a previously reported characterization of the TLGR~\cite{Bobowski:2016}.  

Notice that $\left\vert S_{11,\mathrm{e}}\right\vert$ does not fall below $0.835$ and that this minimum occurs at a frequency of \mbox{$\SI{1.033}{\giga\hertz}$} which is greater than $f_0$.  Furthermore, taking the ratio of the observed $\left\vert S_{11,\mathrm{e}}\right\vert$ resonant frequency and bandwidth yields \mbox{$\sim 300\ll Q_0$}.  All of these observations are indicative of an overcoupled TLGR.  When the best-fit values of $f_0$, $Q_0$, and $M_0^2/R_0$ are substituted into Eqs.~(\ref{eq:Re}) and (\ref{eq:Xe}), the resonator impedance at $f=\SI{1.033}{\giga\hertz}$ is \mbox{$Z_\mathrm{e}=\left(6.1-35.9j\right)\SI{}{\ohm}$} which is a poor match to $Z_0$.  Adjusting the coupling strength to reduce $M_0^2/R_0$ to \SI{8.5}{}\% of the fit value would result in a good match to $Z_0$ at a resonant frequency of \mbox{$f=\SI{1.02935}{\giga\hertz}$}.  In this case, the TLGR would be critically coupled and would have a much sharper resonance.  We emphasize, however, that changing the coupling in this way would adversely affect the strength of the observed resonances in $\left\vert S_{11,\mathrm{f}}\right\vert$.  

\begin{figure*}
\begin{tabular}{lr}
(a)\includegraphics[width=7.7 cm]{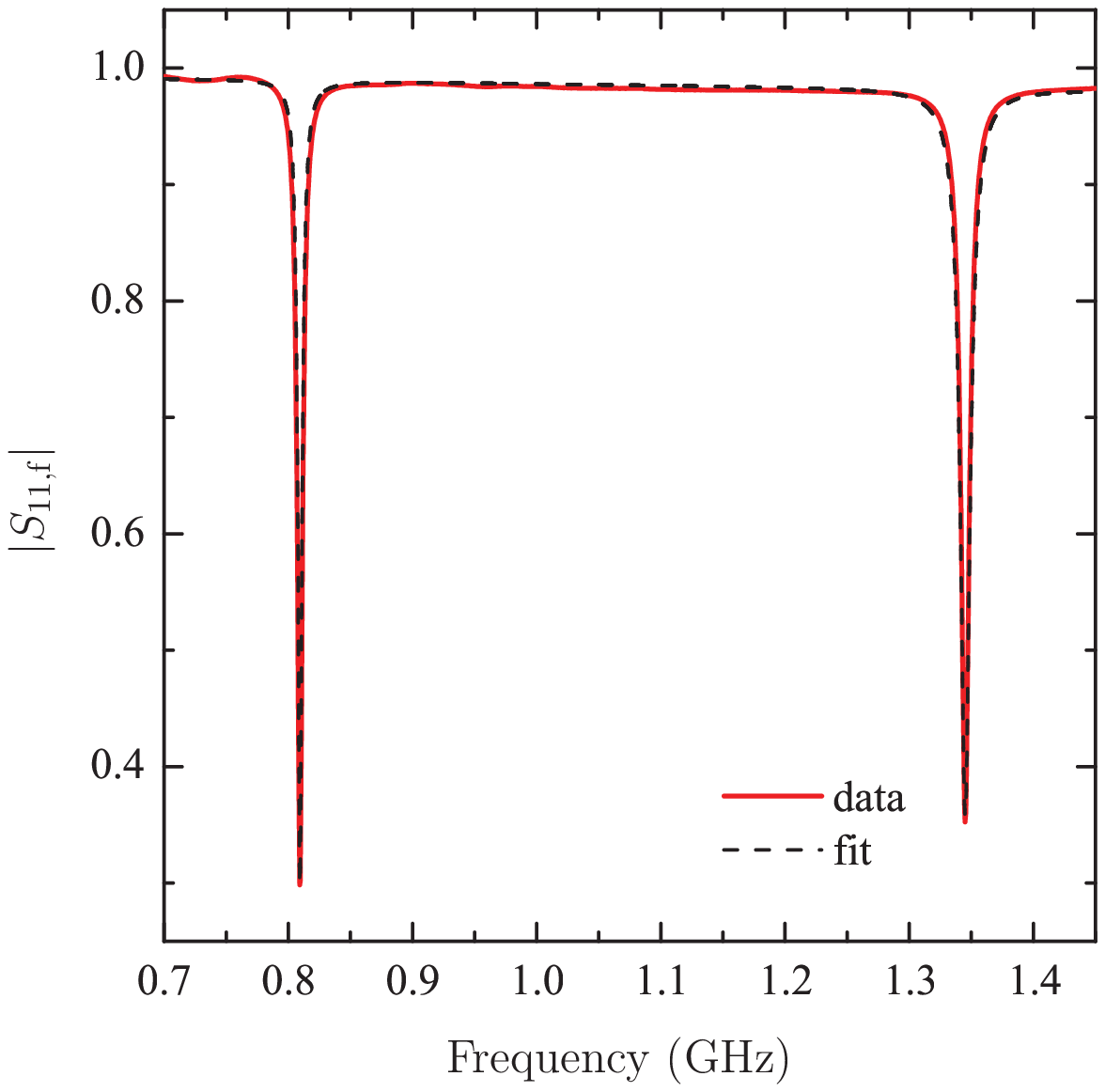}\quad~ & ~\quad (b)\includegraphics[width=7.7 cm]{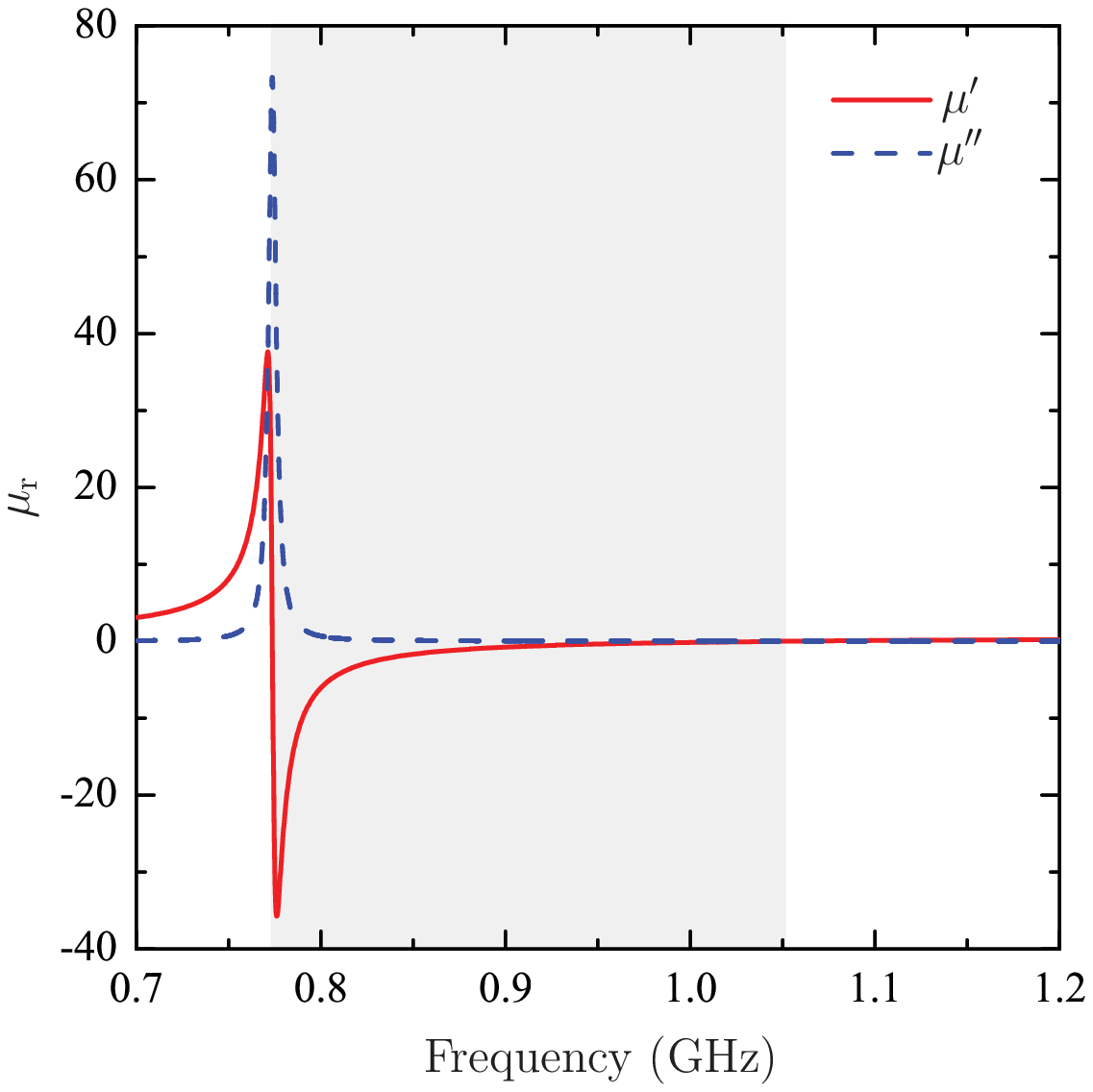}
\end{tabular}
\caption{\label{fig:SRRloadedLGR}(a) Magnitude of the reflection coefficient from an inductively coupled TLGR loaded with an ESRR.  The coupling loop was oriented to achieve the maximum possible signal on the VNA.  The dashed line is a fit to the data.  (b) The real and imaginary components of the ESRR's relative permeability using the parameters extracted from the fit to the data in (a).  Between the frequencies $f_\mathrm{S}$ and $f_\mathrm{p}$ (gray-shaded region), $\mu^\prime$ is negative.}
\end{figure*}

Next, the $\left\vert S_{11,\mathrm{f}}\right\vert$ data were fit to Eq.~(\ref{eq:S11mag}) while using Eqs.~(\ref{eq:Rf}) and (\ref{eq:Xf}) for the real and imaginary components of $Z_\mathrm{f}$.  For these fits, we used the results of Sec.~\ref{sec:loop} for $L_1$ and the results of the empty fit for $f_0$, $Q_0$, and $M_0^2/R_0$.  The remaining fit parameters were the filling factor $\eta$ and the parameters of the ESRR permeability $f_\mathrm{S}$, $f_\mathrm{P}$, and $\gamma_\mathrm{S}/2\pi$.  For this fit, we subtracted $\omega/b^\prime$ from $\left\vert S_{11,\mathrm{f}}\right\vert$ to account for losses in the coupling loop transmission line which increase with increasing frequency.  The fit to the data is shown in Fig.~\ref{fig:SRRloadedLGR}(a).
Although the fit slightly underestimates of the width of the low-frequency resonance and slightly overestimates of that of the high-frequency resonance, it is otherwise very good.  The extracted best-fit parameters were \mbox{$\eta=\SI{0.31\pm 0.05}{}$}, \mbox{$f_\mathrm{S}=\SI{0.77\pm 0.03}{\giga\hertz}$}, \mbox{$f_\mathrm{P}=\SI{1.05367\pm 0.00003}{\giga\hertz}$}, \mbox{$\gamma_\mathrm{S}/2\pi=\SI{4.9\pm 0.4}{\mega\hertz}$}, and \mbox{$b^\prime/2\pi=\SI{75\pm 1}{\giga\hertz}$}.  The frequency dependencies of the both the real and imaginary parts of the ESRR relative permeability determined from these fit parameters and Eqs.~(\ref{eq:mu1}) and (\ref{eq:mu2}) are shown in Fig.~\ref{fig:SRRloadedLGR}(b).  The ESRR is paramagnetic at frequencies below $f_\mathrm{S}$, and $\mu^\prime$ is negative for \mbox{$f_\mathrm{S}\le f\le f_\mathrm{P}$}.

The fit value of the filling factor $\eta$ is approximately \SI{50}{}\% greater than than our estimate of \SI{0.195}{} based on the arc\-length of the ESRR.  We speculate that this enhancement of $\eta$ is due to the fringing magnetic fields that extend beyond the ends of the ESRR.  Using the arc\-length of the ESRR to fix the value of $\eta$ leads to a best-fit value for $f_\mathrm{S}$ that is unrealistically low.  When $\eta$ is left as a free parameter, on the other hand, the fitting routine gives a value of $f_\mathrm{S}$ that is slightly larger than that which was observed when the ESRR was suspended in air (see Fig.~\ref{fig:unshieldedSRR}).  We note, however, that the bore the TLGR is effectively an EM shield surrounding the ESRR and a slight increase in the resonance frequency is, therefore, to be expected~\cite{Hardy:1981}.  From Eq.~(\ref{eq:wP}), \mbox{$f_\mathrm{P}=f_\mathrm{S}/\sqrt{F}$} where $F$ is the fraction of the cross-sectional area of the TLGR's bore that is not occupied by the ESRR.  In our case, we estimate that $F\approx 0.4$ which predicts a magnetic plasma frequency that is \SI{20}{}\% greater than the value found from the fit.  Finally, from Eq.~(\ref{eq:gammaS}), recall that $\gamma_\mathrm{S}/2\pi$ is expected to be given by $f_\mathrm{S}\delta_\mathrm{S}/r$ where $\delta_\mathrm{S}$ is the skin depth of copper when $f=f_\mathrm{S}$.  If $r=\SI{5.0}{\milli\meter}$ is taken to be the average radius of the elliptical ESRR and a copper resistivity of \SI{1.7}{\micro\ohm\centi\meter} is assumed, one calculates \SI{0.36}{\mega\hertz} for $\gamma_\mathrm{S}/2\pi$.  This calculated value is about an order of magnitude less than the fit result.  A greater-than-expected experimental damping constant is perhaps not surprising given the crudeness of the ESRR design and construction.

\section{Experimental Signature of $\bm{\mu^{\prime\prime}<0}$}\label{sec:negMu2}
This section briefly considers the expected $\left\vert S_{11,\mathrm{f}}\right\vert$ response when the bore of a LGR is loaded with a material having $\mu^{\prime\prime}<0$.  In Fig.~3 of Ref.~[\onlinecite{Koschny:2003}], Koschny {\it et al.\@} show an antiresonant frequency dependence for the permeability of a periodic lattice of cut wires.  Specifically, $\mu^\prime$ goes through a resonance, but always remain positive, while $\mu^{\prime\prime}$ is negative at the resonant frequency.  

These authors argue that $\mu^{\prime\prime}<0$ is physically possible provided that the overall energy dissipated in the metamaterial, given by
\begin{equation}
W=\frac{1}{4\pi}\int \omega\left[\varepsilon^{\prime\prime}\left(\omega\right)\left\vert E\left(\omega\right)\right\vert^2+\mu^{\prime\prime}\left(\omega\right)\left\vert H\left(\omega\right)\right\vert^2\right]\,d\omega,
\end{equation}
remains positive.  In this expression $E\left(\omega\right)$ and $H\left(\omega\right)$ are the electric and magnetic fields within the metamaterial, respectively.  Inside the bore of a LGR, however, the electric fields are very weak and the stored energy is predominantly magnetic.  Therefore, for a metamaterial loaded in the bore of a LGR, a negative value of $\mu^{\prime\prime}$ would seem to be problematic.% (unless $\varepsilon^{\prime\prime}$ was simultaneously positive and exceedingly large). 

\begin{figure*}
\begin{tabular}{lr}
(a)\includegraphics[width=7.7 cm]{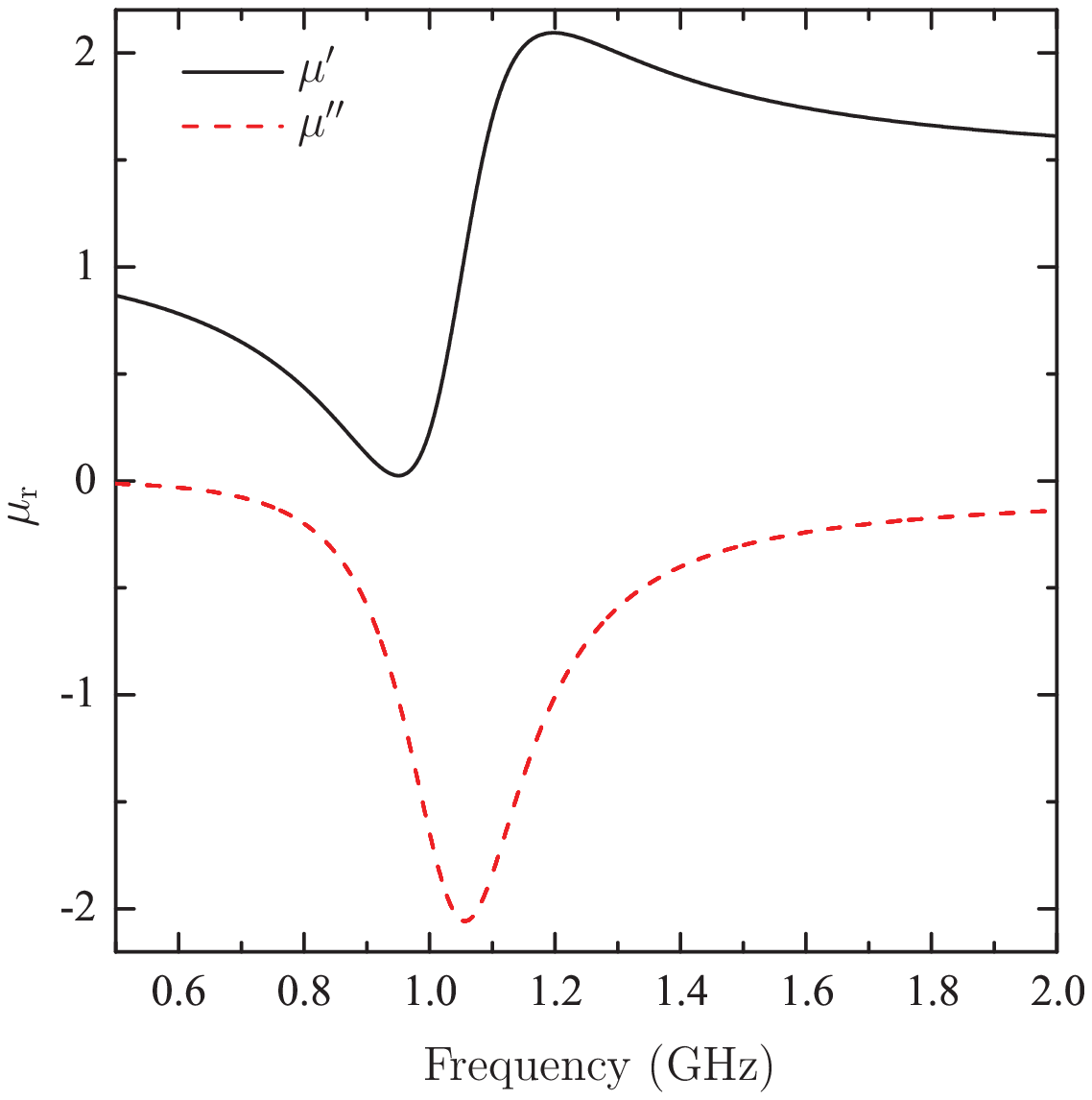}\quad~ & ~\quad (b)\includegraphics[width=7.7 cm]{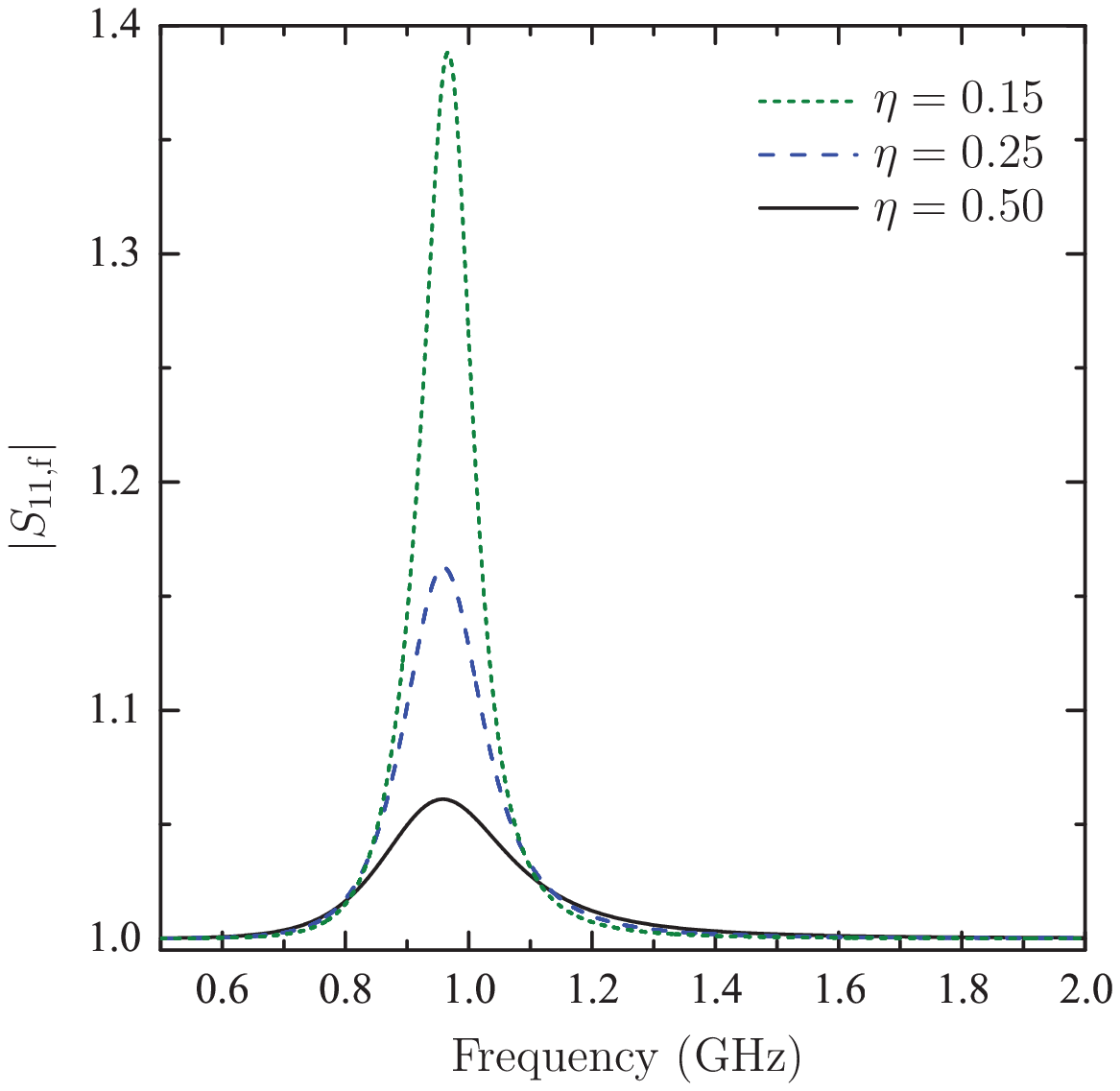}
\end{tabular}
\caption{\label{fig:negMu2}(a) The real and imaginary components of a relative permeability made to mimic that shown for a lattice of cut wires in Fig.~3 of Ref.~[\onlinecite{Koschny:2003}].  (b) The calculated reflection coefficient from an inductively-coupled LGR when its bore is filled with a material having the permeability shown in (a).  The filling factor $\eta$ is varied from $0.15$ to $0.50$.  The calculated response assumes that $f_0=\SI{1.0}{\giga\hertz}$, $Q_0=500$, $L_1=\SI{12}{\nano\henry}$, and \mbox{$M_0^2/R_0=\SI{0.05}{\nano\henry^2/\milli\ohm}$}. A negative $\mu^{\prime\prime}$ results in an unphysical reflection coefficient with $\left\vert S_{11,\mathrm{f}}\right\vert > 1$.}
\end{figure*}

Figure~\ref{fig:negMu2}(a) shows a permeability that has many of the same qualitative features as the cut-wire permeability shown by Koschny {\it et al.\@} in Ref.~[\onlinecite{Koschny:2003}].
It was generated using 
\begin{align}
\mu^\prime &=1+\dfrac{\left[1-\left(\dfrac{\omega_\mathrm{S}}{\omega_\mathrm{P}}\right)^2\right]\left[1-\left(\dfrac{\omega_\mathrm{S}}{\omega}\right)^2\right]}{\left[1-\left(\dfrac{\omega_\mathrm{S}}{\omega}\right)^2\right]^2+\left(\dfrac{\gamma}{\omega}\right)^2},\label{eq:mu1neg}\\
\mu^{\prime\prime} &=\dfrac{-\dfrac{\gamma}{\omega}\left[1-\left(\dfrac{\omega_\mathrm{S}}{\omega_\mathrm{P}}\right)^2\right]}{\left[1-\left(\dfrac{\omega_\mathrm{S}}{\omega}\right)^2\right]^2+\left(\dfrac{\gamma}{\omega}\right)^2},\label{eq:mu2neg}
\end{align}
and the parameters $f_\mathrm{S}=\SI{1.05}{\giga\hertz}$, $f_\mathrm{P}=\SI{1.44}{\giga\hertz}$, and $\gamma_\mathrm{S}/2\pi=\SI{238}{\mega\hertz}$.  In Fig.~\ref{fig:negMu2}(b), we show the calculated $\left\vert S_{11,\mathrm{f}}\right\vert$ that would be expected for a LGR partially filled with a material having the permeability of Fig.~\ref{fig:negMu2}(a).  The reflection coefficient was calculated using Eq.~(\ref{eq:S11mag}) and Eqs.~(\ref{eq:Rf}) and (\ref{eq:Xf}) for the impedance of the partially-filled LGR.  Note, however, that for this calculation the positive root of $\sin\left(\phi/2\right)$ must be chosen in Eq.~(\ref{eq:sin}) since $\ell_2$, as given by Eq.~(\ref{eq:ell2}), would be positive in the $\mu^{\prime\prime}<0$ case.  $\left\vert S_{11,\mathrm{f}}\right\vert$ was calculated for three different filling factors and, as Fig.~\ref{fig:negMu2}(b) shows, the experimental signature of an antiresonant permeability material within the bore of a LGR would be an unphysical reflection coefficient that has a magnitude greater than one.

\section{Summary and Conclusions}\label{sec:summary}
We have proposed a method that can be used to experimentally determine the complex permeability of a 1-D array of SRRs. In the method, the bore of an inductively-coupled LGR is partially filled with the SRR array.  The resulting reflection coefficient $S_{11}$ depends on the detailed frequency dependence of the real and imaginary parts of the array's effective permeability.  Using multi-loop, multi-gap LGRs, the method can be extended to characterize the effective permeability of small 3-D arrays of SRRs.

Using a TLGR, we demonstrated the technique using a single ESRR of length $z$.  Fits to the magnitude of the reflection coefficient produced a reliable set of parameters that completely characterized the effective permeability of the ESRR.  In future work we plan to repeat these measurements using, in place of the ESRR, a 1-D array of $N$ planar SRRs separated by lattice spacing $a$, where \mbox{$N=z/a$}. 

Finally, motivated by reports of metamaterials with effective parameters exhibiting an antiresonant frequency response, we considered a LGR with its bore partially filled with a $\mu^{\prime\prime}<0$ material.  It was shown that, in such a scenario, the reflection coefficient of the inductively-coupled LGR would have a magnitude greater than one.

\begin{acknowledgments}
We wish to acknowledge the support of Thomas Johnson who generously provided access to the Agilent E5061A VNA. \end{acknowledgments}

\nocite{*}
\bibliography{SRRloadedLGR}% Produces the bibliography via BibTeX.

\end{document}